\documentclass[11pt]{article}

\usepackage{amsmath,amssymb,mathtools}
\usepackage{graphicx}
\usepackage{wrapfig}
\usepackage{slashed}
\usepackage{float}
\usepackage{placeins}
\usepackage{longtable}
\usepackage{xcolor}
\usepackage{subcaption}
\usepackage{tikz}
\usetikzlibrary{arrows.meta}
\usepackage{cite}
\usepackage[a4paper,margin=25mm]{geometry}
\usepackage{hyperref} 
\usepackage{braket}
\newcommand{\be}{\begin{equation}}
\newcommand{\ee}{\end{equation}}

\newcommand{\CK}{\mathcal{K}}
\newcommand{\CH}{\mathcal{H}}
\newcommand{\CO}{\mathcal{O}}
\newcommand{\Tr}{\mathrm{Tr}\,}
\newcommand{\Var}{\mathrm{Var}}

\newcommand{\nn}{\nonumber}

\providecommand{\CK}{C_{\mathrm K}}

\providecommand{\Var}{\mathrm{Var}}

\begin{document}

\title{\textbf{Freezing Swampland: A Krylov Complexity Diagnostic for the Weak Gravity Conjecture}}

\author{
\textbf{Amin Faraji Astaneh}$^{a,b}$\footnote{faraji@sharif.ir}
\quad and \quad
\textbf{Reza Ghomi Shurkaie}$^{a}$\footnote{reza.ghomishorkaie@physics.sharif.edu}
}

\date{}
\maketitle

\begin{center}
\emph{$^{a}$ Department of Physics, Sharif University of Technology,\\
P.O. Box 11155-9161, Tehran, Iran}
\end{center}

\begin{center}
\emph{$^{b}$ Research Center for High Energy Physics,\\
Department of Physics, Sharif University of Technology,\\
P.O. Box 11155-9161, Tehran, Iran}
\end{center}

\begin{abstract}
\noindent
We propose a quantum-information-theoretic perspective on the Weak Gravity
Conjecture through the behavior of Krylov spread complexity. For a charged
thermofield double state holographically dual to an AdS
Reissner--Nordstr\"om black hole, we show that the extremal limit on the
black-hole side is accompanied by an effective freezing of Krylov spreading.
In this regime, the return amplitude becomes effectively a pure phase,
and the corresponding Krylov spread is suppressed. We then incorporate
charged matter and study the effect of Schwinger pair production in the
near-horizon region. Within this semiclassical near-extremal analysis, the
frozen behavior is lifted once an allowed discharge channel is opened, and
the dynamics become nontrivial again. This suggests a possible
complexity-based diagnostic of a characteristic physical consequence of the
Weak Gravity Conjecture: the availability of a superextremal charged state
that can mediate discharge. More broadly, our results point to a connection
between the swampland program, black-hole physics, and quantum information
theory.
\end{abstract}

\vskip 7cm

\newpage
\tableofcontents

\section{Introduction}

Two active directions in quantum gravity research are the swampland program
and the interface between black hole physics and quantum information theory.
At first sight, these subjects may appear largely unrelated. Upon closer
inspection, however, both aim to diagnose consistency conditions on quantum
gravity, suggesting that they may be connected in nontrivial ways.

The swampland program seeks to distinguish low-energy effective theories that
admit a consistent ultraviolet completion including gravity from those that
do not, and therefore belong to the swampland rather than the landscape
\cite{Vafa:2005ui,Ooguri:2006in,Brennan:2017rbf,Palti:2019pca}. String theory
provides the best-developed framework in which such questions can be studied
concretely. At the same time, major effort has been devoted to understanding
black holes from the perspective of quantum information theory, motivated in
part by longstanding problems such as the black hole information paradox
\cite{PhysRevD.14.2460,Page:1993wv,Susskind:2005js}.

It is therefore natural to ask whether these two directions are related, and
in particular whether concepts from quantum information theory can provide
useful diagnostics for swampland constraints. If so, one may hope to
reformulate at least some swampland criteria in terms of more general
dynamical or informational properties of quantum-gravitational systems.

In the present work, we focus on one of the best-known swampland criteria,
the Weak Gravity Conjecture (WGC)
\cite{Arkani-Hamed:2006emk,Harlow:2022ich,Montero:2018fns}. In its simplest
form, the WGC states that gravity should be the weakest force. In the
black-hole context, a standard expectation is that sufficiently extremal
charged black holes should not remain absolutely stable when superextremal
charged states are available to mediate discharge.

From the quantum-information side, our main tool will be quantum complexity,
which aims to characterize the growth and spread of quantum states under
time evolution. Among the several notions of complexity that have been
proposed, we focus on Krylov complexity, and in particular on Krylov spread
complexity, which is defined systematically through the Lanczos construction
and the resulting state dynamics in Krylov space
\cite{Krylov1931,Lanczos1950,Parker:2018yvk,Sanchez-Garrido:2024pcy,
Balasubramanian:2022tpr,Nandy:2024evd,Baume:2026jyt,Dymarsky:2021bjq,
Avdoshkin:2022xuw,Alishahiha:2022nhe,FarajiAstaneh:2025thi}. This notion has
proved useful as a diagnostic of quantum dynamics, chaos, and information
spreading, while avoiding the need to invoke a separate holographic
complexity conjecture.

The natural setting in which to bring these themes together is holography,
especially the AdS/CFT correspondence
\cite{Maldacena:1997re,Witten:1998qj,Aharony:1999ti}. Motivated by this
perspective, we study the Krylov complexity of a charged thermofield double
state, holographically dual to an AdS Reissner--Nordstr\"om black hole.
This setup allows us to probe the near-extremal regime and ask how the
approach to extremality is encoded in the spreading dynamics of the dual
state. Related questions have previously been explored using other notions
of holographic complexity, such as the complexity$=$volume and
complexity$=$action proposals
\cite{Stanford:2014jda,Brown:2015bva,Brown:2015lvg,Carmi:2016wjl,
Carmi:2017jqz,Alishahiha:2015rta}, and we comment on these connections
where appropriate.

Our main observation is that, within this framework, the approach to
extremality is accompanied by an effective freezing of Krylov spreading:
the return amplitude becomes increasingly phase-dominated, and the
corresponding spread complexity is strongly suppressed in the extremal
limit. We do not interpret this as a complete dynamical description of
extremal black-hole decay. Rather, we view it as evidence that extremality
is associated with an obstruction to sustained complexity growth in the
dual description.

We then incorporate charged matter and study the opening of a discharge
channel through Schwinger pair production in the near-horizon region. In
this setting, the frozen behavior is replaced by nontrivial Krylov
evolution. Our treatment is semiclassical and restricted to the
near-extremal AdS$_2$ throat, and therefore does not attempt to capture the
full time-dependent, backreacting evolution of the discharging black hole.
Instead, it isolates the instability mechanism and its effect on Krylov
dynamics in a controlled regime.

The resulting picture is not a derivation of the WGC from first principles,
but rather a proposed quantum-information-theoretic diagnostic of one of
its characteristic physical consequences in this holographic setting. To
the best of our knowledge, the specific connection between the absence of
Krylov freezing and the opening of a near-extremal discharge channel has
not been formulated in this way before. More broadly, our results suggest
a program for probing swampland constraints through dynamical complexity
diagnostics.

The paper is organized as follows. In Section~\ref{grand-canonical}, we
formulate Krylov complexity for a charged thermofield double state in the
grand canonical ensemble and relate the return amplitude to the analytically
continued partition function. In Section~\ref{rnads}, we review the charged
AdS Reissner--Nordstr\"om black-hole geometry and its near-extremal
AdS$_2$ throat. In Section~\ref{near-extremal-krylov}, we compute the
near-extremal cumulants and show that Krylov complexity freezes in the
extremal limit. In Section~\ref{decaying_black_hole}, we include charged
matter, derive the Schwinger discharge rate, and show how this channel
unfreezes Krylov dynamics within the semiclassical near-horizon
approximation. We conclude in Section~\ref{conclusion} with comments on
the interpretation of our results and possible future directions. We review
Krylov spread complexity and present several technical details in the
appendices.

\section{Complexity of a system in the grand canonical ensemble}
\label{grand-canonical}

We aim to investigate the balance between mass and charge in a quantum
mechanical system. It is therefore natural to work in the grand canonical
ensemble. In this ensemble, we define the grand-canonical Hamiltonian and
partition function as
\be
K \equiv H - \mu Q \,,
\qquad
Z(\beta,\mu)=\Tr e^{-\beta K}\,,
\ee
where \(\beta\) is the inverse temperature and \(\mu\) is the chemical
potential. We assume that \(H\) and \(Q\) commute, so that they can be
simultaneously diagonalized.

The thermofield double (TFD) state purifies the grand-canonical density
matrix
\be
\rho_{\beta,\mu}=\frac{e^{-\beta K}}{Z(\beta,\mu)}.
\ee
It is defined by
\cite{Takahashi:1996zn,Maldacena:2001kr,Maldacena:2017axo,Cottrell:2018ash,
Hartnoll:2009sz,Doroudiani:2019llj}
\begin{equation}
\ket{\mathrm{TFD}}
=
\frac{1}{\sqrt{Z(\beta,\mu)}}
\sum_n e^{-\frac{\beta}{2}K_n}
\ket{n}_L\ket{n}_R \, ,
\end{equation}
with
$
K_n \equiv E_n-\mu Q_n
$,
where \(\{E_n\}\) and \(\{Q_n\}\) are the eigenvalues of the Hamiltonian
\(H\) and the conserved charge \(Q\), respectively. The labels \(L\) and
\(R\) denote two identical, non-interacting copies of the same quantum
system, with total Hilbert space
\begin{equation}
\mathcal{H}_{\mathrm{tot}}
=
\mathcal{H}_L\otimes \mathcal{H}_R .
\end{equation}
The states \(\ket{n}_L\) and \(\ket{n}_R\) are the corresponding energy and
charge eigenstates in the left and right copies. The TFD state entangles the
two copies such that tracing out one side reproduces the grand-canonical
thermal state on the other.

Taking \(\ket{\mathrm{TFD}}\) as the initial state \(\ket{\psi(0)}\), we now
study its time evolution and the associated return amplitude. Evolving one
copy with the grand-canonical Hamiltonian gives the return amplitude
\be
A(t)
=
\Tr\!\left(\rho_{\beta,\mu}e^{-itK}\right)
=
\frac{Z(\beta+it,\mu)}{Z(\beta,\mu)}\, .
\ee

Two comments are in order. First, one might ask why we evolve with \(K\)
rather than \(H\). The reason is that the TFD state defined above purifies
the grand-canonical density matrix. If we evolved with \(H\), we would
instead probe the different generating function
$
\Tr\!\left(\rho_{\beta,\mu}e^{-itH}\right)
$,
which is not obtained from the analytic continuation \(\beta\to \beta+it\)
of \(Z(\beta,\mu)\) with fixed \(\mu\).

Second, one may ask why we evolve only one copy. This becomes clear when
relating the TFD state to the two-sided eternal RN--AdS black hole. In that
geometry, the global time evolution translates the right boundary forward
and the left boundary backward. Thus, the bulk Killing generator is
\be
K_{\rm global}=K_R-K_L .
\ee
With the above TFD convention, this gives
\be
\bra{\mathrm{TFD}}
e^{-it(K_R-K_L)}
\ket{\mathrm{TFD}}
=
1 .
\ee
Therefore, to obtain a nontrivial return amplitude, we evolve only a single
copy, for instance with \(K_R\). By contrast, evolving both sides forward
with \(K_L+K_R\) gives
\be
\bra{\mathrm{TFD}}
e^{-it(K_L+K_R)}
\ket{\mathrm{TFD}}
=
\frac{Z(\beta+2it,\mu)}{Z(\beta,\mu)} ,
\ee
which corresponds to doubling the imaginary-time shift. Henceforth, $K$ without a subscript denotes the
grand-canonical generator acting on a single copy.

Let us now analyze the return amplitude more systematically. The logarithm
of \(A(t)\) is the cumulant-generating function:
\be
\log A(t)
=
\sum_{n=1}^{\infty}
\frac{(-it)^n}{n!}\,\kappa_n ,
\qquad
\kappa_n
=
(-1)^n
\partial_\beta^n \log Z(\beta,\mu)\Big|_{\mu} .
\ee

Introducing
\[
\delta K \equiv K-\langle K\rangle ,
\]
the first four cumulants are
\begin{align}
\kappa_1 &= \langle K\rangle\, , \nonumber \\
\kappa_2 &= \langle(\delta K)^2\rangle = \Var(K)\, , \nonumber \\
\kappa_3 &= \langle(\delta K)^3\rangle\, , \nonumber \\
\kappa_4 &= \langle(\delta K)^4\rangle
-3\langle(\delta K)^2\rangle^2\, .
\end{align}

In particular, the first and second cumulants are the mean and variance of
the grand-canonical Hamiltonian, respectively.

Using the standard moment--cumulant relations, the first four moments
\(\mu_n=\langle K^n\rangle\) are
\begin{align}
\mu_1 &= \kappa_1, \nn \\
\mu_2 &= \kappa_2+\kappa_1^2, \nn \\
\mu_3 &= \kappa_3+3\kappa_2\kappa_1+\kappa_1^3, \nn \\
\mu_4 &= \kappa_4+4\kappa_3\kappa_1+3\kappa_2^2
+6\kappa_2\kappa_1^2+\kappa_1^4 .
\end{align}

Consequently, the first Lanczos coefficients are
\begin{align}
a_0 &= \kappa_1, \nn \\
b_1^2 &= \kappa_2, \nn \\
a_1 &= \kappa_1+\frac{\kappa_3}{\kappa_2}, \nn \\
b_2^2 &= 2\kappa_2+\frac{\kappa_4}{\kappa_2}
-\frac{\kappa_3^2}{\kappa_2^2}.
\end{align}

and thus the Krylov complexity, up to order $\CO(t^4)$, can be written in
terms of the cumulants as
\begin{equation}
\begin{split}
C_K(t)
&=b_1^2t^2+\frac{1}{12}b_1^2
\left[2b_2^2-4b_1^2-(a_1-a_0)^2\right]t^4+\CO(t^6)\\
&=
\kappa_2 t^2
+
\left(
\frac{1}{6}\kappa_4
-
\frac{1}{4}\frac{\kappa_3^2}{\kappa_2}
\right)t^4
+
\mathcal{O}(t^6).
\end{split}
\end{equation}

Thus, the leading term in the Krylov complexity is determined by the
variance of the grand-canonical Hamiltonian.

The connection to the Weak Gravity Conjecture arises from evaluating these
cumulants for charged black holes, especially near extremality. This allows
us to probe the behavior of Krylov complexity in the extremal limit. This
will be the main objective of the following analysis.

\section{The Weak Gravity Conjecture in the dual geometry}
\label{rnads}

The Weak Gravity Conjecture (WGC) concerns the relationship between mass and
charge in a gravitational theory. It is therefore natural to begin our
study with Einstein--Maxwell theory. In the following subsection, we briefly
review the relevant black hole geometry. This material is included primarily
for completeness; readers interested in the subsequent complexity analysis
may extract the final thermodynamic quantities and proceed directly.

\subsection{Review of charged AdS black hole solutions}

We consider Einstein--Maxwell theory in the presence of a negative
cosmological constant, as our goal is to investigate its holographic dual,
see \cite{Hawking:1982dh,Chamblin:1999hg,Chamblin:1999tk,Witten:1998zw,
Maldacena:2001kr,Hartnoll:2009sz}. The action is given by
\be
S=\int d^{d+1}x\,\sqrt{-g}\left[
\frac{1}{2\kappa^2}\left(R-2\Lambda\right)
-\frac{1}{4g^2}F_{\mu\nu}F^{\mu\nu}
\right],
\ee
where \(\kappa^2=8\pi G_{d+1}\), \(g\) is the gauge coupling, and
\(\Lambda=-\frac{d(d-1)}{2L_{d+1}^2}\) is the negative cosmological constant
with $L_{d+1}$ as the AdS radius.

The Maxwell and Einstein equations are
\be
\nabla_\mu F^{\mu\nu}=0,
\ee
and
\be
G_{\mu\nu}+\Lambda g_{\mu\nu}=\kappa^2 T_{\mu\nu},
\qquad
T_{\mu\nu}=\frac{1}{g^2}\left(
F_{\mu\rho}F_{\nu}{}^{\rho}
-\frac{1}{4}g_{\mu\nu}F_{\rho\sigma}F^{\rho\sigma}
\right).
\ee

To solve these equations, we adopt a static, spherically symmetric ansatz
with a purely electric gauge field:
\be
ds^2=-f(r)\,dt^2+\frac{dr^2}{f(r)}+r^2 d\Omega_{d-1}^2,
\qquad
A=A_t(r)\,dt,
\ee
with the metric function
\be
f(r)=1+\frac{r^2}{L_{d+1}^2}
-\frac{2m}{r^{d-2}}
+\frac{q^2}{r^{2(d-2)}}.
\ee

Solving the Maxwell equation yields the gauge potential
\be
A_t(r)=\mu-\frac{Q g^2}{(d-2)\,\Omega_{d-1}\,r^{d-2}},
\ee
where \(\Omega_{d-1}\) is the volume of the unit \((d-1)\)-sphere and
\(Q = \frac{1}{g^2}\int_{S_\infty} *F\) is the conserved physical charge.
The parameters \(m\) and \(q\) in the metric are related to the ADM mass
\(M\) and the charge \(Q\) via
\be
m=\frac{\kappa^2 M}{(d-1)\Omega_{d-1}},
\qquad
q=\frac{\kappa g Q}
{\sqrt{(d-1)(d-2)}\,\Omega_{d-1}}.
\ee

The horizon radii are the positive roots of \(f(r)=0\). The outer event
horizon is located at \(r_+\), and the Hawking temperature is given by
\be
T=\frac{f'(r_+)}{4\pi}
=\frac{1}{4\pi}\left(
\frac{d-2}{r_+}
+\frac{d\,r_+}{L_{d+1}^2}
-\frac{(d-2)\,q^2}{r_+^{2d-3}}
\right).
\ee

Regularity of the gauge field at the horizon requires \(A_t(r_+)=0\), which
determines
\be
\mu=\frac{Q g^2}
{(d-2)\,\Omega_{d-1}\,r_+^{d-2}}
\quad \implies \quad
q^2 = \gamma \, r_+^{2(d-2)},
\ee
where we have defined the dimensionless parameter
\be
\gamma \equiv
\left(\frac{d-2}{d-1}\right)
\left(\frac{\kappa\mu}{g}\right)^2.
\ee

In terms of \(\mu\), the temperature in the grand-canonical ensemble is
\be
T(r_+;\mu)=\frac{1}{4\pi}
\left[
\frac{(d-2)(1-\gamma)}{r_+}
+\frac{d\,r_+}{L_{d+1}^2}
\right].
\ee

\subsection{Extremality}

The extremal radius \(r_e\) (at fixed \(\mu\)) is defined by
\(T(r_e;\mu)=0\), which satisfies \(f(r_e)=0\) and \(f'(r_e)=0\). In the
near-extremal regime, we hold \(\mu\) fixed while taking \(r_+\to r_e\) and
\(T\to 0\). Expanding \(T(r_+;\mu)\) around \(r_e\) yields
\be
r_+-r_e
\approx
\frac{2\pi L_{d+1}^2}{d}\,T \vert_{\mu}.
\ee

To examine the near-horizon geometry, we introduce the coordinate
\(r=r_+ + u\) with \(|u|\ll r_+\). The metric function expands as
\be
f(r) \approx4\pi T\,u+\frac{1}{2} f''(r_e)\,u^2.
\ee

Let us define
\[
\frac{1}{2}f''(r_e)\equiv \frac{1}{L_2^2},
\]
and introduce the shifted radial coordinate
\be\label{rho}
\rho \equiv u + 2\pi T L_2^2,
\qquad
\rho_0 \equiv 2\pi T L_2^2.
\ee

The metric then takes the form
\be
ds^2 \approx
-\frac{\rho^2-\rho_0^2}{L_2^2}\,dt^2
+\frac{L_2^2}{\rho^2-\rho_0^2}\,d\rho^2
+r_e^2\,d\Omega_{d-1}^2.
\ee

This is the metric of an AdS$_2$ black hole $\times\, S^{d-1}$, which reduces
to AdS$_2 \times S^{d-1}$ in the extremal limit \(T\to 0\). It is now clear
that the constant \(L_2\) is the AdS$_2$ radius.

Near the horizon, the gauge potential is linear in the radial coordinate
\be
A_t \approx E(\rho-\rho_0),
\qquad
E \equiv \frac{(d-2)\mu}{r_e},
\qquad
F_{\rho t}=E.
\label{nearhorizon_Efield}
\ee

After a Wick rotation \(t \to -i\tau\), regularity at
\(\rho=\rho_0\) fixes the Euclidean period \(\tau \sim \tau + \beta\), where
\be
\beta = \frac{2\pi L_2^2}{\rho_0}.
\label{beta_rho0_relation}
\ee

Thus, \(\rho_0\) serves as the parameter characterizing the deviation from
extremality, with \(\rho_0 \to 0\) as \(T \to 0\), as we have already seen
in \eqref{rho}.

Having established the geometric and thermodynamic background of the
charged AdS black hole, we now address the primary focus of this work: the
behavior of Krylov complexity in the extremal regime. From a holographic
perspective, the charged black hole is dual to a charged thermofield double
state in the boundary quantum system. We will investigate how the associated
Krylov spread complexity evolves as the extremal limit is approached and
whether this behavior provides a quantum-information-theoretic signature of
the Weak Gravity Conjecture.

\section{Spread Krylov complexity in the near-extremal limit}
\label{near-extremal-krylov}

Using the AdS/CFT correspondence in the semiclassical approximation, the
grand-canonical partition function is
\begin{equation}
Z(\beta,\mu)\approx
\exp\!\left[-I_E^{\text{on-shell}}(\beta,\mu)\right]
= e^{-\beta\Omega},
\end{equation}
where
\begin{equation}
\Omega = M - TS - \mu Q
\label{logZ_thermo}
\end{equation}
is the grand potential.

The first cumulant is therefore
\begin{equation}
\kappa_1 = -\partial_\beta \log Z = M-\mu Q.
\end{equation}

The second cumulant is
\begin{equation}
\kappa_2 = \partial_\beta^2 \log Z
= -\partial_\beta \kappa_1
= -\partial_\beta(M-\mu Q)\Big|_\mu
= T^2 C_\mu,
\end{equation}
where
\begin{equation}
C_\mu \equiv
T\left(\frac{\partial S}{\partial T}\right)_\mu
\end{equation}
is the heat capacity at fixed chemical potential.

We now consider the near-extremal limit. It is convenient to introduce the
extremal entropy
\begin{equation}
S_e \equiv S(r_e)
=\frac{2\pi}{\kappa^2}\Omega_{d-1}r_e^{d-1}.
\end{equation}

We may then expand the entropy as
\begin{equation}
S(T)=S_e+
\left.\frac{dS}{dr_+}\right|_{r_e}(r_+-r_e)+\cdots
=
S_e+
\left(
\frac{2\pi(d-1)L_{d+1}^2}{d\,r_e}\,S_e
\right)T+\cdots.
\end{equation}

It then follows that
\begin{equation}
C_\mu
=
T\left(\frac{\partial S}{\partial T}\right)_\mu
=
\left(
\frac{2\pi(d-1)L_{d+1}^2}{d\,r_e}\,S_e
\right)T+\cdots.
\end{equation}

Hence, the second cumulant behaves as
\begin{equation}
\kappa_2=\alpha T^3+\mathcal{O}(T^4),
\qquad
\alpha\equiv
\frac{2\pi(d-1)L_{d+1}^2}{d\,r_e}\,S_e.
\end{equation}

One may also rewrite \(\alpha\) entirely in terms of \(S_e\) by substituting
for \(r_e\).

Moreover, from the recursion relation that generates the higher-order
cumulants, one finds the hierarchy
\begin{equation}
\kappa_3 \sim T^4,\qquad
\kappa_4 \sim T^5,\qquad
\text{etc.}
\label{kappascaling}
\end{equation}
while \(\kappa_1\) remains finite in the extremal limit.

Therefore, the Krylov spread complexity takes the form
\begin{equation}
\label{vanishing}
C_K(t)
=
\alpha\,T^3 t^2
+\cdots
\;\xrightarrow{T\to 0}\;
0
\qquad
(\text{for fixed } t,\mu).
\end{equation}

Equivalently, since all higher cumulants satisfy
\(\kappa_{n\ge 2}\to 0\) as \(T\to 0\), the return amplitude reduces to a
pure phase,
\begin{equation}
A(t)=
\exp\!\left(
\sum_{n\ge 1}
\frac{(-it)^n}{n!}\kappa_n
\right)
\;\xrightarrow[T\to 0]{}\;
\exp\!\big(-it\,\kappa_{1,e}\big),
\end{equation}
with no nontrivial broadening generated by the higher cumulants. Here
\(\kappa_{1,e}\) is the extremal value of the first cumulant.

This behavior admits a natural interpretation. The vanishing of $C_K(t)$ in
the extremal limit indicates that the state ceases to spread nontrivially in
the Krylov basis: the dynamics is effectively frozen at the level of
operator growth or state spreading. In the language of the Krylov chain,
hopping away from the initial site becomes increasingly suppressed as
extremality is approached, so the wavefunction remains localized near its
starting point rather than exploring the chain. The extremal limit thus
corresponds to a freezing of motion along the Krylov chain, reflected in the
disappearance of spread complexity. From the holographic viewpoint, this
provides a quantum-information-theoretic signal that extremality is
associated with a highly constrained and, in this sense, non-generic
dynamical regime.

There is a subtle but important point that should be emphasized. In the
black-hole setting, the vanishing of the temperature has a direct geometric
meaning that is central to our analysis. For the RN--AdS black hole,
$T\to0$ corresponds to approaching the extremal geometry, where the two
horizons coalesce and the near-horizon region develops the characteristic
extremal throat. Thus, the zero-temperature limit is not merely a formal
limit of the grand-canonical Gibbs state: in the gravitational system it
has a precise geometric realization as the approach to an extremal black
hole. This geometric interpretation is essential for our purpose, since it
allows us to identify the suppression of thermal Krylov spreading with the
near-extremal-to-extremal limit and subsequently ask whether this
effective freezing persists once charged sectors capable of discharging
the extremal black hole are included.

At the same time, we stress that the vanishing of the higher cumulants of
$K=H-\mu Q$ in the $T\to0$ limit is not, by itself, a specifically
black-hole or WGC effect. It is a general kinematical property of a
grand-canonical Gibbs state when the state becomes concentrated on a
definite eigenspace of $K$. The black-hole-specific content of our
analysis therefore enters through the geometric identification of the
$T\to0$ limit with the near-extremal RN--AdS geometry and its approach to
the extremal configuration, and through the subsequent question of whether
this extremal configuration remains metastable once the charged sectors
responsible for discharge are included. In this sense, the geometric
realization of the $T\to0$ limit provides the physical setting in which the
otherwise kinematical freezing phenomenon becomes relevant to extremal
black-hole stability and WGC-motivated discharge.

This observation may be compared with the results of \cite{Montero}, where
it was found, within the complexity $=$ action proposal, that the extremal
limit can lead to an apparent violation of Lloyd's bound \cite{Lloyd}. The
authors argue, however, that this should not be regarded as a genuine
inconsistency. Their reasoning is that Lloyd's bound relies on assumptions
appropriate to a computational model built from elementary gates that nearly
orthogonalize the quantum state at each step, as in a conventional serial
quantum circuit. Such assumptions are not obviously applicable to
holographic black holes. Their main claim is that, if one adopts the
complexity $=$ action proposal, then large charged AdS black holes are more
naturally described in terms of simple gates, namely gates that induce only
very small rotations of the wavefunction rather than nearly orthogonalizing
it. In that case, the assumptions underlying Lloyd's derivation break down,
and one should therefore not expect the bound to hold in the near-extremal
regime.

Our focus here, however, is on the instability of extremal black holes.
More precisely, we will show that, once the decay process is taken into
account, the system is driven to evolve nontrivially along the Krylov chain.
From this perspective, our goal is to identify a quantum-information-
theoretic manifestation of the Weak Gravity Conjecture.

To close this section, we emphasize that we no longer interpret
Eq.~\eqref{vanishing} as a statement that the exact extremal theory has
identically vanishing complexity for all times. Rather, it is a
semiclassical near-extremal result obtained at fixed time. In particular,
the limits
\begin{equation}
    T\rightarrow0
    \qquad\text{and}\qquad
    t\rightarrow\infty
\end{equation}
need not commute. Schematically,
\begin{equation}
    \lim_{T\to0} C_K(t)\big|_{t\ {\rm fixed}}=0
\end{equation}
is the limit computed in the manuscript, whereas sufficiently late times
may probe effects not captured by the semiclassical near-extremal expansion,
including possible nonperturbative dynamics within the extremal sector.

\section{Complexity of a decaying black hole}
\label{decaying_black_hole}

We investigate whether the freezing of Krylov spreading, and the
corresponding suppression of spread complexity, can serve as a
quantum-information-theoretic diagnostic of extremal black-hole
metastability. We do not assume that the absence of a WGC state is
equivalent to absolute stability, since other discharge mechanisms may be
present. Instead, our objective is to identify the imprint of an explicitly
available charged discharge channel on Krylov dynamics. To this end, we
enlarge the theory by introducing a charged matter sector capable of
coupling to the black-hole degrees of freedom.

In this extended setup, the near-horizon electric field instigates the
production of charged pairs via the Schwinger mechanism
\cite{Schwinger:1951nm}. These decay channels populate previously
inaccessible sectors of the global Hilbert space, describing charged matter
excitations coupled to a partially discharged black hole. In this section,
we demonstrate that this instability provides the physical mechanism
necessary to unfreeze the Krylov dynamics of the coupled
black-hole--matter system.

\subsection{Charged scalar probe in the near-horizon throat}

To diagnose the instability, we introduce a complex scalar field \(\phi\) of
mass \(m_\phi\) and charge \(q_\phi\), governed in Euclidean signature by the
quadratic action
\begin{equation}
I_E[\phi]
=
\int d^{d+1}x\,\sqrt{g}\,
\phi^* \Delta_\phi \phi,
\qquad
\Delta_\phi
=
-\widehat{\nabla}^2+m_\phi^2+\xi R,
\qquad
\widehat{\nabla}_\mu
=
\nabla_\mu-iq_\phi A_\mu.
\label{scalar_action_decay}
\end{equation}

The Gaussian path integral reads
\begin{equation}
\log Z_\phi
=
-\log\det\Delta_\phi
=
\int_0^\infty \frac{ds}{s}\,
\Tr\,e^{-s\Delta_\phi},
\label{proper_time_complex_decay}
\end{equation}
up to the usual local ultraviolet counterterms.

As discussed above, near extremality the geometry develops an
\(\mathrm{AdS}_2\) throat times the horizon sphere,
\begin{equation}
\mathcal{M}_{\rm NH}
\simeq
\mathrm{AdS}_2(L_2)\times S^{d-1}(r_e).
\end{equation}

The scalar curvature of this product geometry is
\begin{equation}
R_{\rm NH}
=
-\frac{2}{L_2^2}
+\frac{(d-1)(d-2)}{r_e^2}.
\end{equation}

We therefore define the effective mass of the $j$-th Kaluza--Klein mode as
\begin{equation}
m_j^2
=
m_\phi^2+\xi R_{\rm NH}
+\frac{j(j+d-2)}{r_e^2}.
\label{jmass}
\end{equation}

The degeneracy of the scalar harmonics on $S^{d-1}$ is
\begin{equation}
d_j
=
\frac{(2j+d-2)(j+d-3)!}{j!(d-2)!}.
\end{equation}

For a constant near-horizon electric field, the relevant local quantity is
\begin{equation}
\omega \equiv |q_\phi E|.
\end{equation}

The imaginary part of the one-loop effective action can then be extracted
from the poles of the proper-time kernel. In the locally constant field
approximation, the leading contribution to the decay rate is
\begin{equation}
\Gamma
\simeq
\ell\,\frac{\omega}{2\pi}
\sum_{j=0}^{\infty}d_j
\exp\!\left(
-\frac{\pi}{\omega}m_j^2
\right),
\label{GammaLCFA}
\end{equation}
where \(\ell\) denotes the regulated proper length of the near-horizon
region.

Retaining only the lowest partial wave gives
\begin{equation}
\Gamma^{(j=0)}
\simeq
\ell\,\frac{\omega}{2\pi}
\exp\!\left(
-\frac{\pi}{\omega}m_0^2
\right),
\qquad
m_0^2
=
m_\phi^2+\xi R_{\rm NH}.
\label{j0}
\end{equation}
This expression makes explicit that the discharge rate is exponentially
suppressed for heavy or weakly charged fields and enhanced by a strong
near-horizon electric field.

The angular-momentum tower on \(S^{d-1}\) can substantially enhance the
discharge rate. For a large black hole, many angular-momentum modes contribute
appreciably. At large \(j\),
\begin{equation}
d_j
\sim
\frac{2}{(d-2)!}j^{d-2},
\qquad
m_j^2
\simeq
m_\phi^2+\frac{j^2}{r_e^2}.
\end{equation}

In the dilute-instanton regime, the sum may be approximated by an integral:
\begin{align}
\Gamma
&\simeq
\ell\frac{\omega}{\pi(d-2)!}
e^{-\pi m_\phi^2/\omega}
\int_0^\infty
dj\,
j^{d-2}
\exp\!\left(
-\frac{\pi j^2}{\omega r_e^2}
\right)
\nonumber\\
&=
\ell\frac{
\Gamma\!\left(\frac{d-1}{2}\right)
}{
2(d-2)!
}
\left(\frac{\omega}{\pi}\right)^{(d+1)/2}
\exp\!\left(
-\frac{\pi}{\omega}m_\phi^2
\right)r_e^{d-1}.
\label{entropy_enhancement_decay}
\end{align}

Thus, the total discharge rate is enhanced by the angular-momentum tower,
or equivalently by the transverse horizon volume. Since the extremal entropy
scales as
\[
S_e\propto r_e^{d-1},
\]
this enhancement may also be viewed as entropic in origin, up to the
appropriate gravitational normalization.

The LCFA captures the local Schwinger pole structure, but not the global
\(\mathrm{AdS}_2\) spectral condition. In the throat, the pair-production
channel is open for supercritical modes satisfying
\begin{equation}
\nu_j^2
=
\omega^2L_2^4-L_2^2m_j^2-\frac14
>0.
\end{equation}

For such modes, the leading semiclassical pair-production factor behaves as
\begin{equation}
\exp\!\left[
-2\pi\bigl(\omega L_2^2-\nu_j\bigr)
\right].
\end{equation}

Accordingly, the exact one-loop effective action acquires an imaginary part
only when at least one charged mode is supercritical. Thus, the LCFA
expression above should be regarded as a local semiclassical estimate of
the discharge rate, while the actual metastability threshold is set by the
global \(\mathrm{AdS}_2\) condition \(\nu_j^2>0\).

Indeed, expanding in the strong-field/weak-curvature regime,
\begin{equation}
\omega^2L_2^4
\gg
L_2^2m_j^2+\frac14,
\end{equation}
one finds
\begin{equation}
\nu_j
\approx
\omega L_2^2
-
\frac{1}{2\omega}
\left(
m_j^2+\frac{1}{4L_2^2}
\right),
\end{equation}
which yields
\begin{equation}
\exp\!\left[
-2\pi\bigl(\omega L_2^2-\nu_j\bigr)
\right]
\simeq
\exp\!\left[
-\frac{\pi}{\omega}
\left(
m_j^2+\frac{1}{4L_2^2}
\right)
\right].
\end{equation}

Thus, the global \(\mathrm{AdS}_2\) exponent reduces at leading order to the
LCFA exponent \(\pi m_j^2/\omega\), with the additional
\(1/(4L_2^2)\) term representing the leading global curvature correction.
The problem of pair production in the AdS$_2$ throat has been investigated
in \cite{Pioline:2005pf}; see also \cite{Cai:2020trh}.

In Appendix~\ref{appC}, we reviewed the main steps of that analysis and
showed how our LCFA result can be recovered from it. However, as argued
above, the quantity most relevant for our purposes is the exponential decay
of the imaginary part of the effective action, which can already be read off
directly from the LCFA. The prefactors may eventually shift the complexity
bound, as we will show, and can be included explicitly whenever needed.

Another point worth highlighting is the precise relation between
$\mathrm{AdS}_2$ supercriticality and the Weak Gravity Conjecture (WGC).
For $m_j^2>0$, the supercriticality condition $\nu_j^2>0$ implies
\begin{equation}
\left(\frac{q_\phi}{m_j}\right)^2
>
\frac{1}{E^2L_2^2}
\left(
1+\frac{1}{4m_j^2L_2^2}
\right),
\label{qm-AdS2}
\end{equation}
which defines the mode-resolved critical charge-to-mass ratio
\begin{equation}
\left(\frac{|q_\phi|}{m_j}\right)_{\rm crit}
=
\frac{1}{|E|L_2}
\sqrt{1+\frac{1}{4m_j^2L_2^2}}.
\label{qmcrit}
\end{equation}
Here, $m_j$ is the effective mass of the $j$-th
Kaluza--Klein mode, as defined in Eq.~\eqref{jmass},
whereas $m_\phi$ denotes the mass of the underlying
charged scalar. Thus, the near-horizon $\mathrm{AdS}_2$
analysis provides a \emph{mode-resolved instability
threshold}. The square-root factor encodes the correction
associated with the $\mathrm{AdS}_2$ spectral threshold,
making the condition more restrictive than its leading
weak-curvature approximation.

For the extremal spherical RN--AdS solution considered here,
the extremality conditions $f(r_e)=f'(r_e)=0$ give
\begin{equation}
\frac{1}{L_2^2}
=
\frac{(d-2)^2}{r_e^2}
+\frac{d(d-1)}{L_{d+1}^2},
\end{equation}
and
\begin{equation}
E^2
=
\frac{g^2}{\kappa^2}
\left[
\frac{(d-1)(d-2)}{r_e^2}
+\frac{d(d-1)}{L_{d+1}^2}
\right].
\end{equation}
Introducing the dimensionless parameter
\begin{equation}
x_e
\equiv
\frac{d(d-1)}{d-2}
\frac{r_e^2}{L_{d+1}^2},
\end{equation}
we obtain the exact relation
\begin{equation}
|E|L_2
=
\frac{g}{\kappa}
\sqrt{\frac{d-1+x_e}{d-2+x_e}}.
\label{eq:EL2-extremal}
\end{equation}
The dependence on $x_e$ smoothly interpolates between
the planar or large-black-hole regime and the
asymptotically flat limit. In particular,
\begin{equation}
|E|L_2
\longrightarrow
\begin{cases}
\displaystyle
\frac{g}{\kappa},
& x_e\to\infty\, ,
\\
\displaystyle
\frac{g}{\kappa}
\sqrt{\frac{d-1}{d-2}},
& x_e\to0\, .
\end{cases}
\end{equation}
Here, the latter limit corresponds to
$L_{d+1}\to\infty$ at fixed $r_e$.

To compare the near-horizon threshold with the standard
WGC, we consider the lowest harmonic, $j=0$, with minimal
curvature coupling, for which $m_0=m_\phi$.
Combining Eqs.~\eqref{qmcrit} and
\eqref{eq:EL2-extremal}, in the asymptotically
flat and weak-curvature regime,
$x_e\to0$ and $m_\phi L_2\gg1$, we obtain
\begin{equation}
\left(\frac{|q_\phi|}{m_\phi}\right)_{\rm crit}
\simeq
\frac{\kappa}{g}
\sqrt{\frac{d-2}{d-1}},
\end{equation}
which coincides with the standard asymptotically flat
extremality bound in our conventions.
It is convenient to introduce the dimensionless
superextremality parameter
\begin{equation}
z_\phi
\equiv
\frac{g|q_\phi|}{\kappa m_\phi}
\sqrt{\frac{d-1}{d-2}},
\label{zphi}
\end{equation}
so that $z_\phi=1$ corresponds to the classical
asymptotically flat extremality bound. The standard
electric WGC requires the existence of at least one
charged state satisfying $z_\phi\geq1$.

For the minimally coupled lowest harmonic, the
$\mathrm{AdS}_2$ supercriticality threshold therefore
approaches the standard superextremality condition
$z_\phi\gtrsim1$ in the combined asymptotically
flat and weak-curvature limit. At finite curvature,
however, the condition $\nu_j^2>0$ imposes a
geometry- and mode-dependent near-horizon instability
criterion. In particular, higher harmonics and
nonminimal curvature couplings modify the effective
mass $m_j$ and hence the corresponding threshold.
The near-horizon instability condition should
therefore not be identified with the WGC bound itself.

Thus, equation \eqref{qm-AdS2} is best understood as
a mode-resolved near-horizon instability criterion
compatible with the WGC, rather than as a derivation
of the WGC or an exact equivalence between the two
conditions. In particular, the absence of a supercritical
mode in the single-scalar sector considered here does
not imply a violation of the WGC. Conversely, failure
to satisfy a given WGC criterion would not, by itself,
establish absolute black-hole stability, since other
discharge channels may be available\footnote{We thank the referee for their careful and constructive comments, which helped clarify this point.}.

We now enlarge the Hilbert space to include the charged sector responsible
for Schwinger discharge,
\begin{equation}
\mathcal{H}_{\rm full}
=
\mathcal{H}_{f}\otimes\mathcal{H}_{\phi},
\end{equation}
and take the initial state to be
\begin{equation}
|\Psi_0\rangle
=
|f\rangle\otimes |0\rangle,
\end{equation}
where $|f\rangle$ denotes the frozen state of the extremal sector, and
$|0\rangle$ is the initial vacuum state of the charged scalar sector.

The important physical question is therefore not simply whether the
equilibrium state freezes as $T\rightarrow0$, but whether this freezing
persists after charged sectors capable of discharging the extremal black
hole are included.

This leads to two logically distinct statements:
\begin{align*}
T\rightarrow0
&\quad\Longrightarrow\quad
C_K^{\rm thermal}(t)\rightarrow0\, ,
\\
\nu_j^2>0
&\quad\Longrightarrow\quad
\text{black-hole discharge}\, .
\end{align*}
The first statement is a kinematical zero-temperature property of the
thermal sector, whereas the second is a dynamical instability criterion
for the particular charged sector under consideration.

The nontrivial part of our proposal is the relation
\begin{equation}
\text{equilibrium freezing}
\quad
\longrightarrow
\quad
\text{discharging of the black hole}
\quad
\longrightarrow
\quad
\text{Krylov unfreezing}.
\end{equation}

For a unitary evolution generated by a Hermitian operator $K_{\rm full}$,
the exact survival amplitude is
\begin{equation}
A_{\rm surv}(t)
\equiv
\langle\Psi_0|
e^{-iK_{\rm full}t}
|\Psi_0\rangle,
\end{equation}
and therefore
\begin{equation}
P_{\rm surv}(t)
\equiv
|A_{\rm surv}(t)|^2
=
1-
(\Delta K_{\rm full})^2t^2
+O(t^4).
\label{eq:response-survival-short}
\end{equation}
Correspondingly, the exact Krylov expansion gives
\begin{equation}
C_K(t)
=
b_1^2t^2+O(t^4).
\end{equation}
The exact and robust statement is instead
\begin{equation}
C_K(t)
=
\sum_{n\geq1}n|\phi_n(t)|^2
\geq
\sum_{n\geq1}|\phi_n(t)|^2
=
1-|\phi_0(t)|^2,
\end{equation}
or
\begin{equation}
\boxed{
C_K(t)\geq1-P_{\rm surv}(t).
}
\label{eq:response-exact-bound}
\end{equation}
The exponential expression
\begin{equation}
P_{\rm surv}(t)\simeq e^{-\Gamma t}
\end{equation}
is now explicitly understood as a semiclassical, coarse-grained decay
law. It follows from the imaginary part of the one-loop effective action
in the appropriate decay regime and is not intended to describe the
universal short-time quantum-Zeno regime. Thus, the exact inequality above
should be used at short times, while in the regime where the exponential
decay law is valid one obtains the parametric estimate
\begin{equation}
C_K(t)
\gtrsim
1-e^{-\Gamma t}.
\label{CK_bound_decay}
\end{equation}
In the simplest situation, in which the decay is
dominated by the leading channel $j=0$, one may identify
$\Gamma\simeq\Gamma^{(0)}$, with
$\Gamma^{(0)}$ given in \eqref{j0}.

We then construct the Lanczos basis associated with the full Krylov
generator $K_{\rm full}$ and the reference state $|\Psi_0\rangle$. The
inequality \eqref{eq:response-exact-bound} shows directly that any
nonzero departure probability implies nonzero Krylov spreading. In this
sense, once a Schwinger discharge channel is available and actually
produces decay, the initial frozen state becomes metastable and the
Krylov dynamics are no longer exactly frozen.

The physical picture is therefore as follows. In the equilibrium
near-extremal black-hole sector, the cumulants $\kappa_{n\ge2}$ vanish as
$T\to0$, and the thermal state becomes localized near the initial site of
the Krylov chain. Once charged matter is included, Schwinger pair production
can populate additional sectors,
\begin{equation}
|\Psi(t)\rangle
=
a_0(t)\,|f;0\rangle
+
\sum_{\lambda}
a_\lambda(t)\,
|f_\lambda;\phi\phi^*\rangle
+\cdots,
\end{equation}
where $|f;0\rangle$ is the initial frozen extremal state, while
$|f_\lambda;\phi\phi^*\rangle$ denotes a state after pair production in the
channel labeled by $\lambda$. The notation $|f_\lambda\rangle$ emphasizes
that the black-hole sector itself changes after emission, so different
discharge channels populate different sectors of the enlarged Hilbert
space.

Schwinger discharge therefore provides a mechanism for unfreezing the
dynamics once the charged black hole is embedded into the full
black-hole--matter Hilbert space. In this sense, the WGC has a suggestive
quantum-information-theoretic interpretation in this setup: a
superextremal charged state that can actually mediate discharge obstructs
permanent freezing in Krylov space. This is a conditional statement about
the specified discharge channel, not a proof that the WGC is equivalent
to the absence of Krylov freezing.

Therefore, in summary,
\begin{equation}
\begin{array}{ll}
t\rightarrow0:
&
P_{\rm surv}(t)
=
1-(\Delta K_{\rm full})^2t^2+\cdots,
\\[2mm]
\text{decay regime}:
&
P_{\rm surv}(t)
\simeq e^{-\Gamma t}.
\end{array}
\end{equation}

The central conclusion concerning unfreezing does not require a nonzero
initial derivative. It follows from the exact inequality
\eqref{eq:response-exact-bound}: whenever the charged sector produces a
nonzero departure probability,
\begin{equation}
1-P_{\rm surv}(t)>0,
\end{equation}
the state necessarily acquires nonzero Krylov spread.

This bound should be contrasted with Lloyd-type bounds, which place an
upper bound on complexity growth in terms of the available energy
\cite{Lloyd}. Our result does not supply such an upper bound; rather, it
shows that once the extremal state is metastable, quantum decay enforces a
nonzero lower bound on the onset of complexity growth. The two statements
are therefore complementary: the available energy limits how fast
complexity can grow, while the Schwinger instability guarantees that, in
the full Hilbert space, the growth cannot remain identically frozen.

It is useful to compare the thermal Krylov time scale with the Schwinger
decay time scale. The thermal Krylov time is defined as
\begin{equation}
\tau_{\rm th}
=
\frac{1}{\sqrt{\kappa_2}}
\simeq
\frac{1}{\sqrt{\alpha}}T^{-3/2},
\label{thermal_decay}
\end{equation}
Thus, $\tau_{\rm th}\to\infty$ as $T\to0$, reflecting the freezing of
ordinary thermal Krylov spreading within the equilibrium black-hole sector.
By contrast, Schwinger discharge introduces the decay time scale
\begin{equation}
\tau_{\rm Sch}
=
\Gamma^{-1},
\end{equation}
defined through the late-time behavior of the vacuum-persistence probability.

We now define the quantity $T_c$ as a parametric crossover scale. It is determined by comparing the thermal
Krylov time with the semiclassical Schwinger decay time:
\begin{equation}
\tau_{\rm th}(T_c)\sim\tau_{\rm Sch}.
\end{equation}
Using
\begin{equation}
\tau_{\rm th}
\simeq
\frac{1}{\sqrt{\alpha}}T^{-3/2},
\qquad
\tau_{\rm Sch}\simeq\Gamma^{-1},
\end{equation}
gives
\begin{equation}
T_c
\sim
\left(
\frac{\Gamma}{\sqrt{\alpha}}
\right)^{2/3}.
\end{equation}
This quantity compares the characteristic time at which ordinary thermal
Krylov spreading becomes appreciable with the characteristic time
associated with the semiclassical loss of survival probability.

The Schwinger rate in the LCFA contains the regulated near-horizon length
$\ell$, as in Eq.~\eqref{GammaLCFA}. Consequently, the numerical value of
$\Gamma$, and hence that of the parametric scale $T_c$, depends on the
precise prescription used to regulate and define the throat. We therefore
do not interpret $T_c$ as a regulator-independent observable or as an
actual crossing point of two Krylov-complexity curves. Rather, it is a
useful parametric estimate of the regime in which Schwinger discharge can
become relevant compared with ordinary thermal Krylov spreading.

\section{Conclusion}
\label{conclusion}

In this paper, we have explored a possible connection between the swampland
program and quantum-information-theoretic aspects of quantum gravity. Our
focus has been on the Weak Gravity Conjecture (WGC) and on whether one of
its characteristic physical implications can be reformulated in terms of
the behavior of quantum complexity. To address this question, we studied
the Krylov spread complexity of a charged thermofield double state in the
grand canonical ensemble, holographically dual to a charged AdS
Reissner--Nordstr\"om black hole.

The first part of our analysis concerned the near-extremal regime of the
dual black-hole geometry. By expressing the cumulants of the return
amplitude in terms of thermodynamic quantities, we showed that all higher
cumulants vanish as the extremal limit is approached, while the first
cumulant remains finite. As a consequence, the return amplitude becomes
effectively a pure phase and the Krylov spread complexity tends to zero.
In the language of Krylov dynamics, the dual state ceases to spread
nontrivially along the Krylov chain: the motion effectively freezes, and
the wavefunction remains localized near its initial site. From the
quantum-information perspective, extremality is therefore associated, in
this setup, with a highly constrained dynamical regime.

The second part of our analysis incorporated the physically expected
discharge channel. By including a charged scalar field in the bulk and
evaluating its one-loop contribution in the near-horizon throat, we showed
that Schwinger pair production generates an imaginary part in the effective
action and hence a nonzero decay rate. This effect removes the exact
freezing of the return amplitude and leads to a nonvanishing lower bound
on the spread complexity, even in the extremal limit. In Krylov-space
language, once the discharge channel is opened, amplitude leaks away from
the initial site and the system resumes nontrivial evolution along the
chain. In this sense, the discharge process unfreezes the dynamics.

We stress that a full treatment of black-hole discharge would require
addressing a genuinely time-dependent and backreacting process, and
therefore lies beyond the scope of the present work; see
\cite{FarajiAstaneh:2026aks} and references therein for studies of Krylov
complexity in time-dependent Hamiltonian systems. In the present analysis,
we restrict attention to the near-extremal AdS$_2$ throat and a
semiclassical approximation, which suffice to isolate the local instability
mechanism and its imprint on Krylov dynamics.

Taken together, our results suggest a conditional quantum-information-
theoretic interpretation of the WGC. When a superextremal charged state is
present and the corresponding near-horizon discharge channel is actually
open, the resulting instability obstructs permanent Krylov freezing in the
enlarged black-hole--matter Hilbert space. The restoration of nontrivial
Krylov spreading can therefore be viewed as a complexity-side manifestation
of this particular physical consequence of the WGC. Our analysis does not
establish that the WGC is equivalent to the absence of Krylov freezing, nor
that WGC violation would imply absolute stability. Rather, it provides a
possible complexity-based diagnostic of an available discharge channel in
the semiclassical RN--AdS setting.

Our proposal should be regarded as a first step rather than a definitive
criterion. It would be interesting to test the same mechanism for other
types of charged matter and alternative decay channels, to investigate
whether similar behavior arises for other notions of complexity, and to
examine whether related ideas can be extended to other swampland conjectures,
such as the distance conjecture. More broadly, it would be worthwhile to
understand whether the freezing or unfreezing of Krylov dynamics can serve
as a systematic diagnostic of quantum-gravity consistency. We hope that the
perspective developed here will help motivate further work at the interface
of swampland ideas, black-hole physics, and quantum information theory.

\section*{Acknowledgements}

We thank H. Arfaei for reading the draft and for insightful discussions.

\newpage
\appendix

\section{A brief review of Krylov complexity}
\label{app}

As with other notions of quantum complexity, \emph{Krylov complexity}
(also called \emph{spread complexity} for states) aims to quantify how
complicated a state becomes under time evolution, in terms of how broadly
it spreads over an appropriate set of elementary building blocks. In this
framework, the building blocks are generated by repeatedly acting with the
Hamiltonian on an initial state and then orthonormalizing the resulting
vectors.

Starting from an initial state
$\ket{\psi(0)}\equiv\ket{\psi_0}\in\CH$, the time-evolved state is
\begin{equation}
\ket{\psi(t)}
=
e^{-iHt}\ket{\psi_0}
=
\sum_{n=0}^{\infty}
\frac{(-it)^n}{n!}\ket{\psi_n},
\qquad
\ket{\psi_n}\equiv H^n\ket{\psi_0}.
\end{equation}

The resulting vectors must be orthonormalized (e.g.\ by the Gram--Schmidt
procedure). In the present context, this orthonormalization is implemented
via the Lanczos algorithm, which produces an orthonormal basis
$\{\ket{k_n}\}$ spanning the Krylov subspace
$\mathcal{K}\subset\mathcal{H}$ of dimension $d_\CK$. These basis vectors
play the role of the elementary building blocks for the time-evolved state.

Setting $\ket{k_0}\equiv\ket{\psi_0}$, the Lanczos recursion is
\begin{equation}
\ket{k_{n+1}}
=
\frac{\ket{\widetilde k_{n+1}}}
{\|\widetilde k_{n+1}\|},
\qquad
\text{where}
\qquad
\ket{\widetilde k_{n+1}}
=
(H-a_n)\ket{k_n}-b_n\ket{k_{n-1}},
\end{equation}
with $b_0=0$. Here $\|\cdot\|$ denotes the norm, and
\begin{equation}
a_n\equiv\bra{k_n}H\ket{k_n},
\qquad
b_{n+1}\equiv\|\widetilde k_{n+1}\|,
\end{equation}
are the Lanczos coefficients. These coefficients determine the effective
tridiagonal representation of $H$ in the Krylov basis and control the
dynamics of amplitudes in Krylov space.

Expanding the time-evolved state in the Krylov basis,
\begin{equation}
\ket{\psi(t)}
=
\sum_{n=0}^{d_\CK-1}\varphi_n(t)\ket{k_n},
\end{equation}
the Schr\"odinger equation implies a nearest-neighbor recursion for the
amplitudes:
\begin{equation}
i\,\dot{\varphi}_n(t)
=
a_n\varphi_n(t)
+b_{n+1}\varphi_{n+1}(t)
+b_n\varphi_{n-1}(t),
\end{equation}
with initial condition $\varphi_n(0)=\delta_{n0}$.

The return amplitude is
\begin{equation}
A(t)
\equiv
\bra{\psi_0}e^{-iHt}\ket{\psi_0}
=
\varphi_0(t).
\end{equation}

Defining the moments
\begin{equation}
\mu_n
\equiv
\bra{\psi_0}H^n\ket{\psi_0}
=
i^n\frac{d^nA(t)}{dt^n}\bigg|_{t=0},
\end{equation}
one can compute the Lanczos coefficients algorithmically. For instance,
the first few moment relations are
\begin{align}
\mu_0 &= 1,\nonumber\\
\mu_1 &= a_0,\nonumber\\
\mu_2 &= a_0^2+b_1^2,\nonumber\\
\mu_3 &= a_0^3+(2a_0+a_1)b_1^2,\nonumber\\
\mu_4 &=a_0^4+
(3a_0^2+2a_1a_0+a_1^2+b_2^2)b_1^2+b_1^4.
\end{align}

The first Krylov amplitudes can then be expanded at early times. Up to
\(O(t^3)\), \(\varphi_1(t)\) is
\begin{equation}
\varphi_1(t)
=
-ib_1t
-
\frac{1}{2}b_1(a_0+a_1)t^2
+
\frac{i}{6}b_1
\left(
a_0^2+a_0a_1+a_1^2+b_1^2+b_2^2
\right)t^3
+
\mathcal{O}(t^4),
\end{equation}
while up to \(O(t^2)\), \(\varphi_2(t)\) is
\begin{equation}
\varphi_2(t)
=
-\frac{1}{2}b_1b_2t^2
+
\mathcal{O}(t^3).
\end{equation}

Finally, the Krylov complexity is defined as the average Krylov index with
respect to the probability distribution
$p_n(t)\equiv|\varphi_n(t)|^2$:
\begin{equation}
C_K(t)
\equiv
\langle n\rangle_t
=
\sum_{n=0}^{d_{\mathcal K}-1}
n\,|\varphi_n(t)|^2.
\end{equation}

Equivalently, $C_K(t)$ measures how far the wavefunction has spread along
the effective one-dimensional Krylov chain.

\section{Derivation of the Charged Scalar Heat Kernel}
\label{appB}

The heat kernel, or proper-time kernel, for the $j$-th Kaluza--Klein mode is
defined as
\begin{equation}
K_j(s;x,x')
=
\langle x|e^{-s\Delta_j}|x'\rangle,
\end{equation}
where, for a minimally coupled charged scalar with Kaluza--Klein mass
$m_j$, the Euclidean differential operator is
\begin{equation}
\Delta_j
=
-\widehat{\nabla}^2+m_j^2.
\end{equation}

Here
\begin{equation}
\widehat{\nabla}_\mu
=
\partial_\mu-iq_\phi A_\mu
\end{equation}
is the gauge-covariant derivative. Since the mass term commutes with the
covariant Laplacian, we can factor it out:
\begin{equation}
e^{-s\Delta_j}
=
e^{s\widehat{\nabla}^2}e^{-sm_j^2}.
\end{equation}

Therefore, we first evaluate the massless part of the heat kernel in a
constant Euclidean background field. In two Euclidean dimensions, with
coordinates
\begin{equation}
x^\mu=(x^0,x^1),
\end{equation}
the covariant derivatives satisfy
\begin{equation}
[\widehat{\nabla}_0,\widehat{\nabla}_1]
=
-iq_\phi F^E_{01},
\end{equation}
where $F^E_{01}$ is the Euclidean field strength. For a physical Minkowski
electric field $E$, the Euclidean field strength is obtained by analytic
continuation,
\begin{equation}
F^E_{01}=iE.
\end{equation}

To compute the Euclidean heat kernel, choose the Landau gauge
\begin{equation}
A_0=-F^E_{01}x^1,
\qquad
A_1=0.
\end{equation}

Then
\begin{equation}
\widehat{\nabla}_0
=
\partial_0+iq_\phi F^E_{01}x^1,
\qquad
\widehat{\nabla}_1
=
\partial_1.
\end{equation}

Since the background is translationally invariant in the $x^0$ direction,
we Fourier transform with respect to $x^0$:
\begin{equation}
\partial_0\to ik.
\end{equation}

The operator $-\widehat{\nabla}^2$ then becomes
\begin{equation}
-\widehat{\nabla}^2
=
-\partial_1^2
+
\left(k+q_\phi F^E_{01}x^1\right)^2.
\end{equation}

For a real Euclidean background, define the Euclidean oscillator frequency
\begin{equation}
\omega_E\equiv|q_\phi F^E_{01}|.
\end{equation}

After shifting the coordinate as
\begin{equation}
y
=
x^1+\frac{k}{q_\phi F^E_{01}},
\end{equation}
the operator takes the harmonic-oscillator form
\begin{equation}
H_{\rm osc}
=
-\frac{d^2}{dy^2}
+\omega_E^2y^2.
\end{equation}

Mehler's formula gives the diagonal heat kernel
\cite{calin_heat_2011,Kleinert2009}
\begin{equation}
\langle y|e^{-sH_{\rm osc}}|y\rangle
=
\sqrt{
\frac{\omega_E}
{2\pi\sinh(2\omega_Es)}
}
\exp\left[
-\omega_Ey^2\tanh(\omega_Es)
\right].
\end{equation}

The coincident heat kernel per unit volume is obtained by integrating over
the conserved momentum $k$:
\begin{align}
K_{E,j}(s;x,x)
&=
e^{-sm_j^2}
\int_{-\infty}^{\infty}
\frac{dk}{2\pi}
\sqrt{
\frac{\omega_E}
{2\pi\sinh(2\omega_Es)}
}
\exp\left[
-\omega_E
\left(
x^1+\frac{k}{q_\phi F^E_{01}}
\right)^2
\tanh(\omega_Es)
\right]
\nonumber\\
&=
e^{-sm_j^2}
\sqrt{
\frac{\omega_E}
{2\pi\sinh(2\omega_Es)}
}
\frac{|q_\phi F^E_{01}|}{2\pi}
\sqrt{
\frac{\pi}
{\omega_E\tanh(\omega_Es)}
}
\nonumber\\
&=
\frac{\omega_E}
{4\pi\sinh(\omega_Es)}
e^{-sm_j^2},
\end{align}
where we have included the dependence on the $j$-th level mass in this
step. This expression is independent of the spacetime point $x$, as
expected for a constant background field. Any coordinate dependence
appearing in an intermediate gauge choice is a gauge artifact and
disappears from the gauge-invariant coincident heat-kernel density.

The free-field limit is recovered smoothly:
\begin{equation}
\lim_{\omega_E\to0}
\left[
\frac{1}{4\pi s}
\frac{\omega_Es}{\sinh(\omega_Es)}
e^{-sm_j^2}
\right]
=
\frac{1}{4\pi s}e^{-sm_j^2}.
\end{equation}

We now analytically continue back to a physical Minkowski electric field.
Since
\begin{equation}
F^E_{01}=iE,
\end{equation}
the Euclidean frequency analytically continues as
\begin{equation}
\omega_E\to i\omega,
\qquad
\omega\equiv|q_\phi E|,
\end{equation}
which gives
\begin{equation}
\frac{\omega_Es}{\sinh(\omega_Es)}
\to
\frac{\omega s}{\sin(\omega s)}.
\end{equation}

Therefore, the exact proper-time kernel for the $j$-th charged scalar mode
in a constant electric background is
\begin{equation}
K_j(s;x,x)
=
\frac{1}{4\pi s}
\frac{\omega s}{\sin(\omega s)}
e^{-sm_j^2},
\qquad
\omega=|q_\phi E|.
\end{equation}

This completes the derivation.

\section{Pair production in the AdS$_2$ throat}
\label{appC}

In this appendix, we review pair production in a globally $\mathrm{AdS}_2$
throat, following \cite{Pioline:2005pf}.

For a spinless charged scalar in $\mathrm{AdS}_2$, the Euclidean one-loop
effective action admits the proper-time representation
\begin{equation}
W_E
=
-\int_0^\infty\frac{ds}{s}
\int_0^\infty d\nu\,
\rho_{\mathcal E}(\nu)\,
e^{-s\left(
\mu_j^2-\mathcal E^2+\nu^2+\frac14
\right)},
\label{ads2_propertime}
\end{equation}
where $\nu$ is the continuous $\mathrm{AdS}_2$ spectral parameter and
$\rho_{\mathcal E}(\nu)$ is the exact electric spectral density,
\begin{equation}
\rho_{\mathcal E}(\nu)
=
\frac{V_{\rm reg}}{2\pi}
\frac{\nu\sinh(2\pi\nu)}
{\cosh(2\pi\nu)+\cosh(2\pi\mathcal E)}.
\label{ads2_density}
\end{equation}

Here $V_{\rm reg}$ denotes the regulated $\mathrm{AdS}_2$ volume, and we
use the dimensionless throat variables
\begin{equation}
\mathcal E
\equiv L_2^2\omega
=
L_2^2|q_\phi E|,
\qquad
\mu_j\equiv L_2m_j,
\label{throat_dimensionless}
\end{equation}
with $L_2$ the $\mathrm{AdS}_2$ radius, $\omega\equiv|q_\phi E|$ the
physical electric coupling frequency introduced in the main text, and
$m_j$ the effective mass of the $j$-th Kaluza--Klein mode,
\eqref{jmass}.

The exact spectral threshold is controlled by
\begin{equation}
\Delta_j
\equiv
\sqrt{
\mathcal E^2-\mu_j^2-\frac14
}
=
\sqrt{
L_2^4\omega^2-L_2^2m_j^2-\frac14
},
\label{deltaj_appendix}
\end{equation}
so the pair-production channel is open only when
\begin{equation}
L_2^4\omega^2
>
L_2^2m_j^2+\frac14.
\label{ads2_threshold}
\end{equation}

This is the global $\mathrm{AdS}_2$ threshold condition for the $j$-th
mode.

When $\Delta_j^2>0$, the proper-time integral develops an unstable
contribution associated with $0<\nu<\Delta_j$. Expanding the spectral
density in the standard series form,
\begin{equation}
\rho_{\mathcal E}(\nu)
=
\frac{V_{\rm reg}\nu}{2\pi}
\sum_{n=1}^{\infty}
(-1)^{n-1}
\left[
e^{-2\pi(\mathcal E-\nu)n}
-
e^{-2\pi(\mathcal E+\nu)n}
\right],
\label{ads2_density_expansion}
\end{equation}
and evaluating the unstable contribution by analytic continuation, one
obtains
\begin{equation}
\operatorname{Im}(W_E)_j
=
\frac{V_{\rm reg}}{4\pi^2}
\sum_{n=1}^{\infty}
\frac{(-1)^n}{n^2}
\left(1-2\pi n\Delta_j\right)
e^{-2\pi n(\mathcal E-\Delta_j)}.
\label{ads2_series}
\end{equation}

This series can be summed in closed form:
\begin{equation}
\operatorname{Im}(W_E)_j
=
\frac{V_{\rm reg}}{4\pi^2}
\left[
\operatorname{Li}_2
\!\left(
-e^{-2\pi(\mathcal E-\Delta_j)}
\right)
-
2\pi\Delta_j\,
\operatorname{Li}_1
\!\left(
-e^{-2\pi(\mathcal E-\Delta_j)}
\right)
\right].
\label{ads2_polylog}
\end{equation}

The local constant-field approximation (LCFA) corresponds to the regime in
which the relevant formation scale of the quantum process is much smaller
than the curvature radius of the throat. Equivalently, the electric field
and the induced tunneling dynamics are effectively constant over the region
that contributes to pair production. In our dimensionless variables this
is the large-$\mathcal E$ limit,
\begin{equation}
\mathcal E=L_2^2\omega\gg1,
\label{lcfalimit}
\end{equation}
which may be realized either by taking $L_2\to\infty$ at fixed $\omega$ or
by considering a sufficiently strong field at fixed $L_2$. In this regime,
curvature effects are parametrically suppressed, and the throat locally
approaches flat space.

Specifically, in the LCFA regime $\mathcal E\gg1$, we expand the threshold
parameter as
\begin{equation}
\Delta_j
=
\sqrt{\mathcal E^2-\mu_j^2-\frac14}
\approx
\mathcal E
-\frac{1}{2\mathcal E}
\left(
\mu_j^2+\frac14
\right),
\end{equation}
so that
\begin{equation}
2\pi n(\mathcal E-\Delta_j)
\approx
\frac{\pi n}{\mathcal E}
\left(
\mu_j^2+\frac14
\right)
\approx
\frac{n\pi m_j^2}{\omega}
+\mathcal O
\left(
\frac{1}{L_2^2\omega}
\right),
\end{equation}
where the term proportional to $1/(L_2^2\omega)$ represents a global
curvature correction that vanishes in the flat-space limit
($L_2\to\infty$). Likewise, the prefactor behaves as
\begin{equation}
1-2\pi n\Delta_j
\approx
-2\pi n\mathcal E.
\end{equation}

Substituting these term-by-term approximations into the spectral sum
\eqref{ads2_series} yields
\begin{align}
\operatorname{Im}(W_E)_j
&\approx
\frac{V_{\rm reg}}{4\pi^2}
\sum_{n=1}^{\infty}
\frac{(-1)^n}{n^2}
\left(-2\pi n\mathcal E\right)
\exp\!\left(
-\frac{n\pi}{\omega}m_j^2
\right)
\nonumber\\
&=
\left(
\frac{V_{\rm reg}}{2\pi}\mathcal E
\right)
\sum_{n=1}^{\infty}
\frac{(-1)^{n+1}}{n}
\exp\!\left(
-\frac{n\pi}{\omega}m_j^2
\right).
\end{align}

Identifying the regulated throat volume with the local flat-space volume
via
\begin{equation}
\frac{V_{\rm reg}}{2\pi}\mathcal E
\longrightarrow
\frac{t\,\ell}{4\pi}\omega,
\end{equation}
and summing over all Kaluza--Klein modes $j$ with degeneracies $d_j$, we
obtain
\begin{align}
\operatorname{Im}W(t)
&\approx
\frac{t\,\ell}{4\pi}\omega
\sum_{j=0}^{\infty}d_j
\sum_{n=1}^{\infty}
\frac{(-1)^{n+1}}{n}
\left[
\exp\!\left(
-\frac{\pi}{\omega}m_j^2
\right)
\right]^n
\nonumber\\
&=
\frac{t\,\ell}{4\pi}\omega
\sum_{j=0}^{\infty}d_j
\log\left[
1+
\exp\!\left(
-\frac{\pi}{\omega}m_j^2
\right)
\right].
\end{align}

which successfully matches the expected flat-space limit
\eqref{GammaLCFA}.


\end{document}